\documentclass[manuscript]{aastex}

\shorttitle{Resonantly damped propagating kink waves in the solar corona}
\shortauthors{G. Verth, M. Goossens \& J. Terradas}

\begin{document}


\title{Observational evidence of resonantly damped \\propagating kink waves in the solar corona}


\author{G. Verth and M. Goossens}
\affil{Centrum voor Plasma Astrofysica and and Leuven Mathematical Modeling and
Computational Science Centre, KU Leuven, Celestijnenlaan 200B, 3001 Heverlee, Belgium}
\email{gary.verth@wis.kuleuven.be}
\email{marcel.goossens@wis.kuleuven.be}

\author{J. Terradas}
\affil{Departament de F\'\i sica, Universitat de les Illes Balears, E-07122 Palma de Mallorca, Spain}
\email{jaume.terradas@uib.es}

\begin{abstract}
In this Letter we establish clear evidence for the resonant absorption damping mechanism by analyzing observational data from the novel Coronal Multi-Channel Polarimeter (CoMP). This instrument has established that in the solar corona there are ubiquitous propagating low amplitude ($\approx$1 km s$^{-1}$) Alfv\'{e}nic waves with a wide range of frequencies. Realistically interpreting these waves as the kink mode from magnetohydrodynamic (MHD) wave theory, they should exhibit a frequency dependent damping length due to resonant absorption, governed by the TGV relation showing that transversal plasma inhomogeneity in coronal magnetic flux tubes causes them to act as natural low-pass filters. It is found that observed frequency dependence on damping length (up to about 8 mHz) can be explained by the kink wave interpretation and furthermore, the spatially averaged equilibrium parameter describing the length scale of transverse plasma density inhomogeneity over a system of coronal loops is consistent with the range of values estimated from TRACE observations of standing kink modes.
\end{abstract}


\maketitle

\section{Introduction}
\label{intro}
Recently, there has been much interest generated by the observation of ubiquitous propagating Alfv\'{e}nic waves in the solar corona detected by \citet{tomczetal07} using the innovative Coronal Multi-Channel Polarimeter (CoMP) instrument. The Alfv\'{e}nic properties of these waves are undeniable since they have a phase speed of about 1 Mm s$^{-1}$, the velocity components are perpendicular to the direction of magnetic field lines and they exhibit only very weak intensity fluctuations, suggesting an almost incompressible wave mode. Detected in Doppler velocity, they have a low amplitude of approximately 1 km s$^{-1}$ peak to peak corresponding to a loop displacement amplitude of only 48 km and peak power at a period of about 5 minutes. The small spatial extent of the displacement explains why these waves could not be detected previously by the Transition Region and Coronal Explorer (TRACE) since it has a resolution of about 730 km. However low amplitude, about 2 km s$^{-1}$ peak to peak, Alfv\'{e}nic waves in coronal loops, also with a peak power period of about 5 minutes, have been detected before in Doppler velocity by the EUV Imaging Spectrometer (EIS) instrument on Hinode, e.g., \citet{erdtar08} and \citet{vandetal08}. The consistent observed properties suggest that the two instruments have detected the same type of propagating coronal wave. \citet{erdtar08} and \citet{vandetal08} interpreted these transversal coronal waves as the kink mode from magnetohydrodynamic (MHD) wave theory \citep[see][for a detailed discussion of this wave mode]{goossens09}.
\section{Kink waves in coronal loops}
\label{Kink waves}
When the first EUV post-flare transversal coronal loop oscillations were directly imaged by TRACE \citep[][]{aschetal99, nakaetal99}, it was clear that the simple model of Alfv\'{e}n waves originally derived by Alfv\'{e}n (1942) based on an infinite homogeneous plasma in planar geometry could not accurately describe the oscillatory properties of coronal loops, which are finite closed magnetic flux tubes with inhomogeneous plasma structure. Coronal observations show the EUV intensity of oscillating loops is greater than the surrounding medium, indicating the internal plasma density is greater inside the loop than in the ambient plasma. Essentially, the simple homogeneous plasma model by Alfv\'{e}n would not permit us to attempt magnetoseismology of the solar atmosphere to determine its fine structure \citep[see e.g.,][for review]{ban07} and would not allow us to estimate the contribution of wave heating to the coronal plasma, since the most likely mechanisms proposed thus far require plasma inhomogeneity in the direction transverse to the direction of the magnetic field, e.g., phase mixing and resonant absorption \citep[see e.g.,][for review of various proposed damping mechanisms]{asch04}. Due to the observed plasma inhomogeneity of coronal loops theorists concluded that the transversal oscillations detected with TRACE were best described as standing kink waves from MHD theory. This mode is a bulk transversal oscillation of magnetic flux tubes that have different internal and external Alfv\'{e}n speed equilibria, primarily due to density variation in the direction transverse to the direction magnetic field \citep[e.g.,][]{edrob83}. Since spectroscopic EUV line ratios can be implemented to determine coronal density using the CHIANTI database \citep[e.g.,][]{landi06}, \citet{erdtar08} and \citet{vandetal08} exploited the spectral capabilities of EIS to determine that the coronal loops in which they detected low amplitude Alfv\'{e}nic waves had greater internal plasma density than external, naturally leading them to the kink mode interpretation, in agreement with the previous consensus regarding transverse waves in coronal loops observed by TRACE.

Notably, the observed coronal loop standing kink waves observed with TRACE are heavily attenuated in $1-4$ fundamental periods \citep[see e.g.,][]{aschetal03} and the most likely physical mechanism for this proposed thus far is resonant absorption caused by naturally occurring gradients in the transversal Alfv\'{e}n speed of coronal loops \citep[see e.g.,][for review]{goossens08}. Although resonant absorption is still a robust damping mechanism in more complex geometries, e.g., multi-thread loop structures \citep{terradas08}, as an illustration of the basic process we assume a coronal loop to be a cylindrical magnetic flux tube. Then the kink phase speed in the long wavelength limit for an axially homogeneous magnetic cylinder with differing internal and external plasma density and constant magnetic field strength is
\begin{equation}
v_{\mathrm{ph}}=B\sqrt{\frac{2}{\mu(\rho_{\mathrm{i}}+\rho_{\mathrm{e}})}},
\label{kspeed}
\end{equation}
where $\rho_{\mathrm{i}}$ and $\rho_\mathrm{{e}}$ are the internal and external plasma density, $B$ magnetic field strength and $\mu$ magnetic permeability. Equation~(\ref{kspeed}) shows that the kink speed is intermediate between the external and internal Alfv\'{e}n speeds. Regarding kink waves in coronal loops, it is observed that $\rho_{\mathrm{i}} > \rho_{\mathrm{e}}$, so naturally occurring density gradients between the internal and external plasma will cause the local Alfv\'{e}n frequencies in this intermediate layer to match the global kink frequency at some magnetic surface, thus causing a resonance. This generates azimuthal Alfv\'{e}nic motions in the resonant surface which grow in amplitude, damping the global kink mode as energy is transferred to this localized mode. For kink waves excited by a broadband disturbance there will be many resonances within a loop and smaller length scales will be created due to phase mixing, causing a cascade of energy to smaller scales where dissipation becomes more efficient.

In the follow up paper by \citet{tomcmac09} more detailed analysis of the same CoMP data set showed that propagating waves traveling a larger distance suffered greater damping. This inspired \citet{pascoe10} and \citet{terrarretal10} to model propagating kink waves damped by the process of resonant absorption. Most relevant to the present study, \citet{terrarretal10} derived simple analytical expressions for the damping length of kink waves, showing damping length is a monotonically decreasing function of frequency, the TGV relation. This has a consequence that solar waveguides with transverse inhomogeneity, e.g., coronal loops, will act as low pass filters for propagating kink waves. It is the main purpose of this Letter to investigate if the propagating Alfv\'{e}nic waves observed by \citet{tomczetal07}, realistically interpreted as the kink mode from MHD wave theory exhibit this frequency dependent damping.

\begin{figure}\centering
\includegraphics[width=8cm,height=4cm]{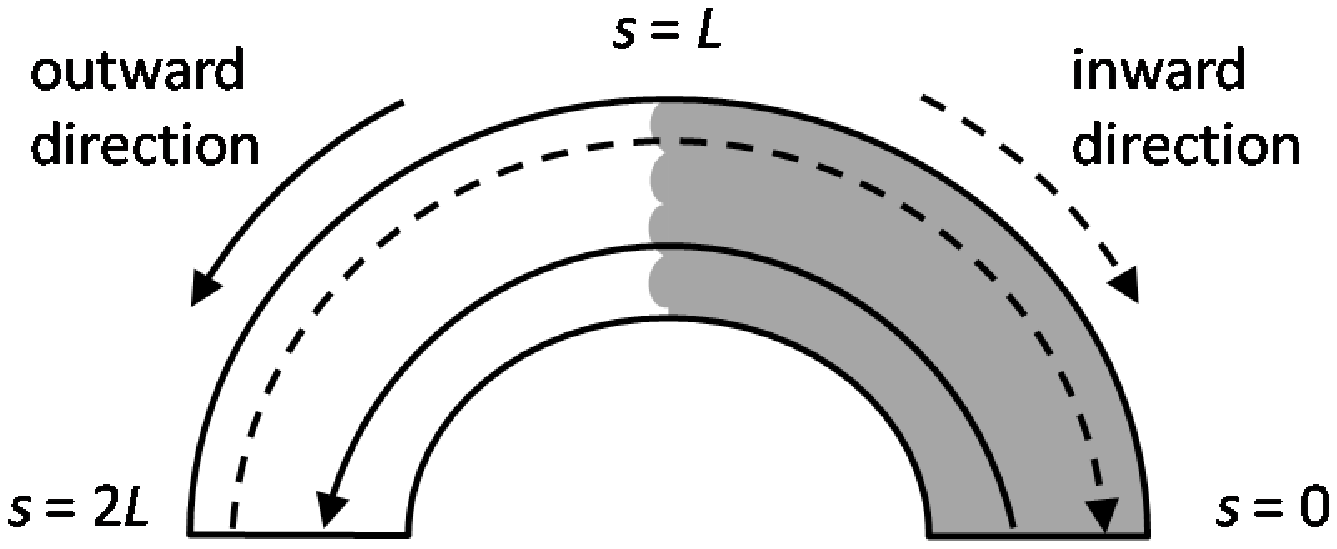}
\caption{Simple cartoon of the observed semi-circular annulus geometry of the coronal loop system. The integrated wave paths analyzed by \citet{tomcmac09} (shaded in grey) are approximately only half the length along the total loop system ($L\approx 250$~Mm) since the data in this segment had the best SNR. The direction of outward and inward wave propagation is shown by the solid and dashed lines respectively.}
\label{fluxtube}
\end{figure}
\section{Theoretical model}
\label{theory}
Traditionally, since standing kink waves in coronal loops were observed using TRACE, the measurement of damping times was of primary interest, see e.g., \citet{aschetal03} and \citet{arregui07}. However, for propagating kink waves it is more appropriate to study the damping length $L_{\mathrm{D}}$, with the expression for wave amplitude given by
\begin{equation}
A(s)=A_{\mathrm{0}} \exp\left(-\frac{s}{L_{\mathrm{D}}}\right),
\label{amp}
\end{equation}
where $s$ is the distance along the waveguide and $A_0$ is initial amplitude. From \citet{tomczetal07}, the transverse scale of propagating disturbances has an upper limit of about 9~Mm and wavelengths are $\lambda \gtrapprox 180$~ Mm, so assuming that the average waveguide observed with CoMP is a fluxtube of radius $R \lessapprox 4.5$ Mm, kink waves are in the long wavelength regime where $R /\lambda\ll 1$. In this limit it was shown by \citet{terrarretal10} that the TGV relation is simply given by
\begin{equation}
L_{\mathrm{D}}=v_{\mathrm{ph}} \, \xi_{\mathrm{E}} \frac{1}{f},
\label{TGV}
\end{equation}
  where $f$ is the frequency defined by $f=v_{\mathrm{ph}}/\lambda$ and $\xi_{\mathrm{E}}$ is an equilibrium parameter dependent on length scale of the density inhomogeneity. Equation (\ref{TGV}) demonstrates that $L_{\mathrm{D}}$ is inversely proportional to $f$, i.e., the rate of damping per unit length will be greater for higher frequency waves than low frequency waves. The parameter $\xi_{\mathrm{E}}$ can be calculated precisely for a chosen equilibrium model, e.g., if we choose a thin inhomogeneous boundary layer with a continuous sinusoidal profile decreasing between $\rho_{\mathrm{i}}$ and $\rho_{\mathrm{e}}$ then
\begin{equation}
\xi_{\mathrm E}=\frac{2}{\pi}\frac{R}{l}\frac{\rho_{\mathrm i}+\rho_{\mathrm
e}}{\rho_{\mathrm i}-\rho_{\mathrm e}},
\label{xidef}
\end{equation}
where $l$ is the thickness of the boundary layer \citep[see e.g.,][]{goossensetal92,rudrob02, goossens02}. Hence Equations (\ref{TGV}) and (\ref{xidef}) demonstrate that the efficiency of damping due to resonant absorption also depends on the thickness of the non-uniform layer and the steepness of the transverse gradient in density.

From Figure~1 (Panel A) in \citet{tomcmac09}, it can be seen that the integrated wave path in the coronal loop system is off limb, with \citet{tomcmac09} assuming the path is in the plane of sky. The approximate semi-circular annular geometry of the this wave path is shown in Figure~\ref{fluxtube}, with the direction of outward and inward propagating waves denoted by arrows. We show the analyzed wave paths by the shaded grey region, which consists of many waves traveling along different coronal loop structures, integrated by \citet{tomcmac09} to improve the Signal to Noise Ratio (SNR), c.f., Figure 4 (Panels B and C) in \citet{tomcmac09}. Although \citet{tomcmac09} did not integrate the entire loop lengths, i.e., from the outward footpoint ($s=0$) to the inward footpoint ($s=2L$), the important point is that in the grey region, inward waves have traveled further and therefore the damping will be greater relative to the outgoing waves.

Let the initial outward at $s=0$ and inward power at $s=2L$ for frequency $f$ be denoted by $P_{\mathrm{out}}(f)$ and $P_{\mathrm{in}}(f)$ respectively. Then since $P(f)\propto A^2(f)$, by Equations~(\ref{amp}) and (\ref{TGV}) the spatially averaged outward power in the shaded grey region $s\in[0,L]$ for frequency $f$ is
\begin{equation}
\left<P(f)\right>_{\mathrm{out}}=\frac{1}{L}\int_0^{L}P_{\mathrm{out}} (f)\exp\left (-\frac{2f}{v_{\mathrm{ph}} \xi_{\mathrm{E}}}s\right)ds
\label{po}
\end{equation}
and for the inward power is
\begin{equation}
\left<P(f)\right>_{\mathrm{in}}=\frac{1}{L}\int_{L}^{2L}P_{\mathrm{in}} (f)\exp\left(-\frac{2f}{v_{\mathrm{ph}} \xi_{\mathrm E}}s\right)ds.
\label{pi}
\end{equation}
 Note that for Equation~(\ref{pi}) integrating the power in the grey region from $s=L$ to $s=0$ for the inward wave is equivalent to integrating from $s=L$ to $s=2L$ for the outward wave. Defining $\left< P(f) \right>_{\mathrm{ratio}}=\left<P(f)\right>_{\mathrm{out}}/ \left<P(f)\right>_{\mathrm{in}}$, it can be shown by Equations~(\ref{po}) and (\ref{pi}) that
\begin{equation}
\left<P(f)\right>_{\mathrm{ratio}}=\frac{P_{\mathrm{out}}(f)}{P_{\mathrm{in}}(f)}\exp\left(\frac{2 L}{v_{\mathrm{ph}}\xi_{\mathrm{E}}}f\right).
\label{p_f}
\label{ratio}
\end{equation}
In the next Section we shall do an exponential least squares fit using Equation~(\ref{ratio}) with CoMP data from the same observation by \citet{tomcmac09} in the frequency range $0~-~4$~mHz where the SNR is strongest.
\label{conclusions}
\begin{figure}\centering
\includegraphics[width=10.0cm,height=5.5cm]{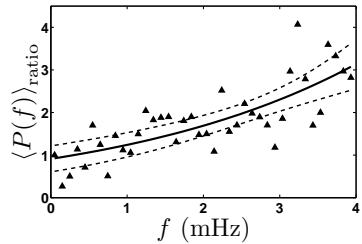}
\caption{Power ratio against frequency best fit using Equation~(\ref{ratio}) with CoMP data. Fixed parameters are $v_{\mathrm{ph}}=0.6$ Mm s$^{-1}$ and $L=250$ Mm. Best estimate parameters are $P_{\mathrm{out}}/P_{\mathrm{in}}=0.91$ and $\xi_{\mathrm{E}}=2.69$ (solid line). The $95\%$ confidence intervals for the simultaneous functional bounds are shown by the dashed lines. Note that in the CoMP data SNR decreases with increasing frequency.}
\label{power}
\end{figure}
\section{Least squares fit to data}
\label{data}
In Figure \ref{power} it can be seen that the CoMP data (see triangles) shows a clear trend of $\left<P(f)\right>_{\mathrm{ratio}}$ increasing with frequency $f$, i.e., that higher frequency waves are damped more than their low frequency counterparts. We add the caveat that in the CoMP data, this trend continues to about 8 mHz, then there is a turn over in the ratio $\left<P(f)\right>_{\mathrm{ratio}}$. At the present time it is unclear if this is simply due to background noise or is the result of some other physical mechanism/s. Understanding this high frequency trend should be the focus of a future study. However, there is a strong level of confidence in the trend of $\left<P(f)\right>_{\mathrm{ratio}}$ in the range $0~-~4$~mHz, which can be explained by the theory of propagating kink wave resonant damping. To illustrate this we implement a least squares fit of the power ratio function given in Equation (\ref{ratio}) to the CoMP data in this range. From the estimates of \citet{tomcmac09} our fixed parameters are $v_{\mathrm{ph}}=0.6$~Mm~s$^{-1}$ and path length $L=250$~Mm for both outward and inward waves, leaving the free parameters as $P_{\mathrm{out}}/P_{\mathrm{in}}$ and $\xi_{\mathrm{E}}$. It is found that the best estimates are $P_{\mathrm{out}}/P_{\mathrm{in}}=0.91$ and $\xi_{\mathrm{E}}=2.69$ (see solid line in Figure~2) with $95\%$ confidence bounds $P_{\mathrm{out}}/P_{\mathrm{in}}=0.67-1.89$ and $\xi_{\mathrm{E}}=1.15-3.49$. Although the wave paths analysed by \citet{tomcmac09} are integrated over many coronal loop structures, the spatially averaged estimate of $\xi_{\mathrm{E}}$ is consistent with the damped standing kink waves in individual coronal loops observed by TRACE where $\xi_{\mathrm{E}}\approx 1-4$ \citep{aschetal03}. It is interesting to note that the average half length of loops estimated with TRACE data is about $L=110$~Mm due to its restricted field of view. The longer value of $L=250$ Mm observed with CoMP suggests that the rate of damping is independent of loop length, i.e., long loops have similar transversal length scale density inhomogeneities as short loops. This has also been independently confirmed recently by \citet{verwichte10} in the study of a long loop of $L=345$ Mm using combined SoHO, TRACE and STEREO observations, where they found $\xi_{\mathrm{E}}=1.48$ for a post-CME standing kink wave. Thus, the analysis of observational cases so far, suggest they are all within the valid regime of the thin tube approximation of resonantly damped kink waves.

Another interesting feature shown in Figure \ref{power} is the estimated least squares value of $P_{\mathrm{out}}/P_{\mathrm{in}}\approx 1$, suggesting that power generated at both the outward and inward footpoints is approximately equal. The $95\%$ confidence intervals for the simultaneous functional bounds, i.e., calculated with all predictor values, are also shown in Figure \ref{power} by the dashed lines. It can be seen that there is a trend of decreasing confidence for $f\gtrapprox 3$ mHz, most likely due to the fact that the SNR for the CoMP data decreases with increasing frequency.

\begin{figure}\centering
\includegraphics[width=10.0cm,height=5.5cm]{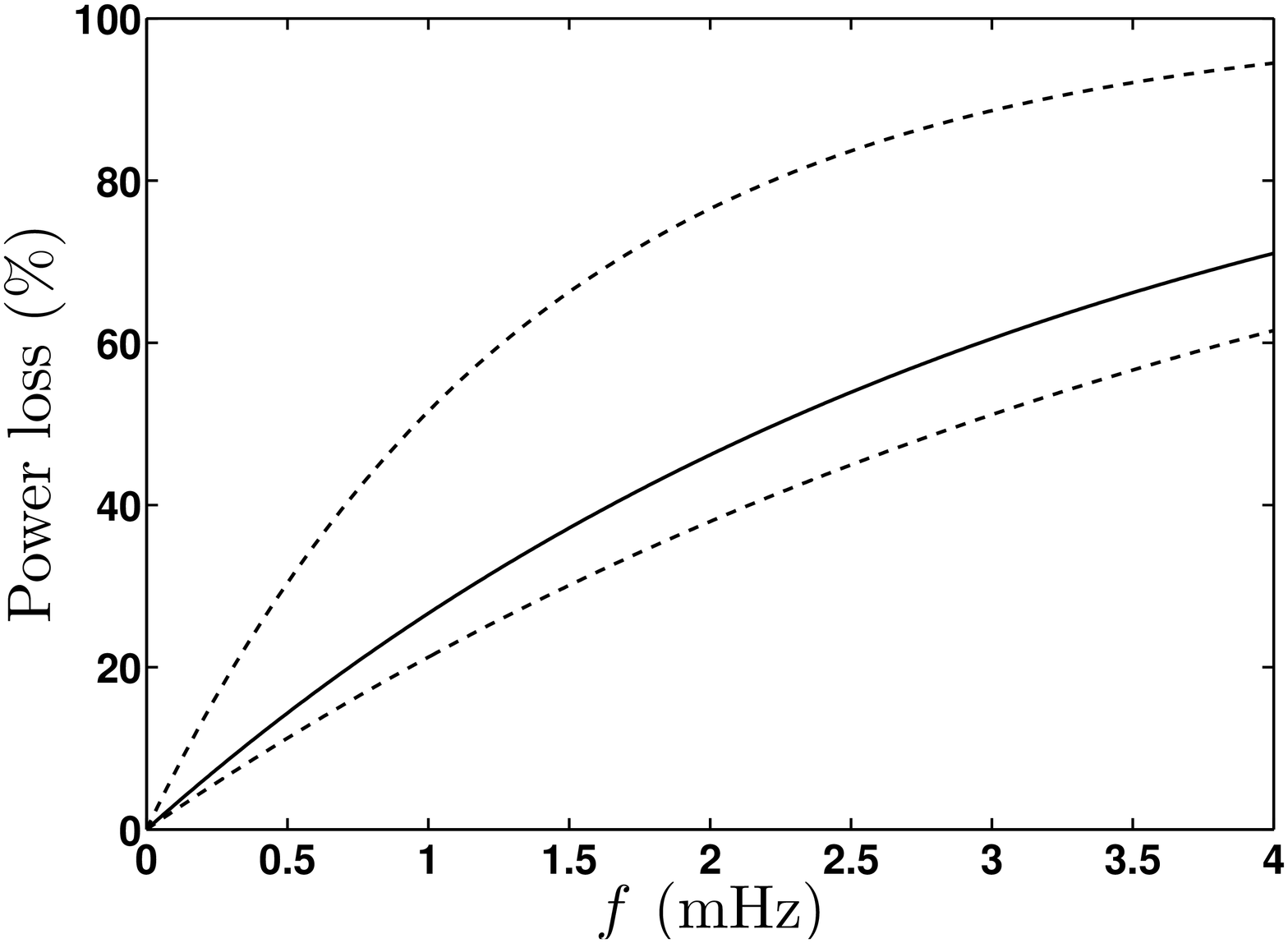}
\caption{Percentage power loss against frequency of a kink wave after one travel time between the two loop footpoints $s=0$ and $s=2L$, a distance of 500 Mm. The dashed lines indicate the percentage power loss up to the 95\% confidence bounds of estimated parameter $\xi_{\mathrm{E}}$.}
\label{ploss}
\end{figure}

From estimates of $L$, $v_{\mathrm{ph}}$ and $\xi_{\mathrm{E}}$, using Equations (\ref{amp}) and (\ref{TGV}) we can calculate the percentage power loss as a function of frequency for the observed propagating kink waves after one travel time between the two loop footpoints, $s=0$ and $s=2L$ (see Figure \ref{ploss}). The kink waves with $f \gtrapprox 2.5$ mHz lose at least 50\% of their power and as explained previously in Section \ref{Kink waves}, this broadband frequency power is converted into Alfv\'{e}nic azimuthal motions in many resonant surfaces.
\section{Conclusions}
In this Letter we established more evidence for the damping mechanism of resonant absorption by analyzing observational data from the CoMP. Crucially, this instrument has established in the solar corona there are ubiquitous propagating low amplitude ($\approx$1 km s$^{-1}$) Alfv\'{e}nic waves with a wide range of frequencies. Realistically interpreting these propagating waves as kink modes, it was predicted they should exhibit a frequency dependent damping length due to resonant absorption, governed by the TGV relation which shows that naturally occurring transversal plasma inhomogeneity in coronal magnetic flux tubes causes them to act as low-pass filters.
It was found the observed frequency dependence on damping length can be explained by the kink wave interpretation, at least up to  about $f=8$~mHz, and furthermore, the spatially averaged equilibrium parameter describing the length scale of transverse plasma density inhomogeneity over a system of coronal loops is consistent with the range of values estimated from TRACE observations of standing kink modes. Due to its restricted field of view, full loops observed with TRACE are shorter than the average loop length of the coronal loop system  observed with CoMP. The similar inferred transversal plasma inhomogeneity length scales, suggest this parameter is independent of loop length. Importantly, it was also found that the estimated least squares value of the ratio of initial outward and inward power was approximately unity, suggesting the kink wave power generated at both footpoints in the lower atmosphere was almost equal.

Previously much work has been undertaken to study the properties of the well known standing kink modes in post-flare coronal loops. The results of this Letter now open the way to investigating the wave contribution to coronal heating by exploiting the latest high spatial/temporal resolution observations of propagating kink waves modes. An accurate estimate of coronal kink wave damping rates due to resonant absorption is crucial in order to quantify the amount of this wave energy that can contribute to plasma heating. Broadband frequency disturbances propagating along transversally inhomogeneous coronal flux tubes will generate many resonances that will phase mix, causing a cascade of energy to smaller scales where dissipation becomes more efficient, thereby creating the necessary conditions for plasma heating. Unlike the larger amplitude standing kink mode coronal loop oscillations which need relatively high energy but rare excitation events like flares or CMEs, small amplitude propagating kink waves are ubiquitous. This makes them an ideal candidate for study with coronal observing instruments of sufficient Doppler velocity resolution such as CoMP, Hinode and the planned Advanced Technology Solar Telescope (ATST).
\begin{acknowledgements}{}
The authors thank S. Tomczyk for his help and expertise regarding the CoMP data. J.T. acknowledges the Universitat de les Illes Balears
for a postdoctoral position and the funding provided under projects
AYA2006-07637 (Spanish Ministerio de Educaci\'on y Ciencia). M.G. and G.V.
acknowledge support from K.U.Leuven via GOA/2009-009. M.G. also expresses gratitude to both the FWO Vlaanderen and UIB.
 \end{acknowledgements}

\end{document}